# Adaptive Noise Cancellation Using Deep Cerebellar Model Articulation Controller

Yu Tsao, *Member*, *IEEE*, Hao-Chun Chu, Shih-Wei Lan, Shih-Hau Fang, *Senior Member*, *IEEE*, Junghsi Lee*, and Chih-Min Lin, *Fellow*, *IEEE*

**Abstract**—This paper proposes a deep cerebellar model articulation controller (DCMAC) for adaptive noise cancellation (ANC). We expand upon the conventional CMAC by stacking single-layer CMAC models into multiple layers to form a DCMAC model and derive a modified backpropagation training algorithm to learn the DCMAC parameters. Compared with conventional CMAC, the DCMAC can characterize nonlinear transformations more effectively because of its deep structure. Experimental results confirm that the proposed DCMAC model outperforms the CMAC in terms of residual noise in an ANC task, showing that DCMAC provides enhanced modeling capability based on channel characteristics.

## I. INTRODUCTION

The goal of an adaptive noise cancellation (ANC) system is to remove noise components from signals. In ANC systems, linear filters are widely used for their simple structure and satisfactory performance in general conditions, where least mean square (LMS) [1] and normalized LMS [2] are two effective criteria to estimate the filter parameters. However, when the system has a nonlinear and complex response, a linear filter may not provide optimal performance. Accordingly, some nonlinear adaptive filters have been developed. Notable examples include the unscented Kalman filter [3, 4] and the Volterra filter [5, 6]. Meanwhile, cerebellar model articulation controller (CMAC), a feed forward neural network model, has been used as a complex piecewise linear filter [7, 8]. Experimental results showed that CMAC provided satisfactory performance in terms of mean squared error (MSE) for nonlinear systems [9, 10].

A CMAC model is a partially connected perceptron-like associative memory network [11]. Owing to its peculiar structure, it overcomes fast growing problems and learning difficulties when the amount of training data is limited as compared to other neural networks [8, 12, 13]. Moreover, because of its simple computation and good generalization capability, the CMAC model has been widely used to control complex dynamical systems [14], nonlinear systems [9, 10], robot manipulators [15], and multi-input multi-output (MIMO) control [16, 147].

Yu Tsao is with the Research Center for Information Technology Innovation, Academia Sinica, Taipei, Taiwan (corresponding author to provide phone: 886227872390; email: yu.tsao@citi.sinica.edu.tw).
Hao-Chun Chu and Shih-Wei Lan, Shih-Hau Fang, Junghsi Lee, and Chih-Min Lin are with the Department of Electrical Engineering, Yuan Ze University, Taoyuan, Taiwan. (e-mail: {david4633221, qooqoo6308} @gmail.com; {shfang, eejlee, cml}@saturn.yzu.edu.tw).

More recently, deep learning has become a part of many state-of-the-art systems, particularly computer vision [18-20] and speech recognition [21-23]. Numerous studies indicate that by stacking several shallow structures into a single deep structure, the overall system could achieve better data representation and, thus, more effectively deal with nonlinear and high complexity tasks. Successful examples include stacking denoising autoencoders [20], stacking sparse coding [24], multilayer nonnegative matrix factorization [24], and deep neural networks [26, 27]. In this study, we propose a deep CMAC (DCMAC) framework that stacks several single-layered CMACs. We also derive a modified backpropagation algorithm to train the DCMAC model. Experimental results on ANC tasks show that the DCMAC provides better results than conventional CMAC in terms of MSE scores.

## II. PROPOSED ALGORITHM

### 2.1 System Overview

Figure 1 shows the block diagram of a typical ANC system containing two microphones, one external and the other internal. The external microphone receives the noise source signal $n(k)$, while the internal one receives the noisy signal $v(k)$. The noisy signal is a mixture of the signal of interest $s(k)$ and the damage noise signal $z(k)$. Therefore, $v(k) = s(k) + z(k)$, where $z(k)$ is generated by passing the noise signal $n(k)$ through an unknown channel $F(\cdot)$. The relation between the noise signal $n(k)$ and damage noise $z(k)$ is from [28]. The ANC system computes a filter, $\hat{F}(\cdot)$, which transforms $n(k)$ to $y(k)$, so that the final output, $(v(k) - y(k))$, is close to the signal of interest, $s(k)$. The parameters in $\hat{F}(\cdot)$ are updated by minimizing the MSE.

Recently, the concept of deep learning has garnered great attention. Inspired by deep learning, we propose a DCMAC framework, which stacks several layers of the single-layered

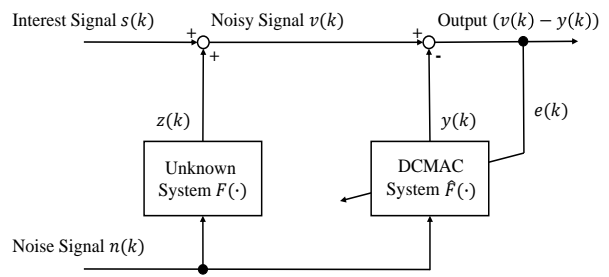

Figure 1. Block diagram of the proposed ANC system.

CMAC, to construct the filter $\hat{F}(\cdot)$, as indicated in Fig. 1. Fig. 2 shows the architecture of the DCMAC, which is composed of a plurality of CMAC layers. The A, R, and W in Fig. 2 denote the association memory space, receptive field space, and weight memory space, respectively, in a CMAC model. In the next section we will detail these three spaces. The output of the first layer CMAC is treated as the input for the next CMAC layer. The derived $\hat{F}(\cdot)$, as modeled by the DCMAC, can better characterize the signals by using multiple nonlinear processing layers, and thus achieve an improved noise cancellation performance.

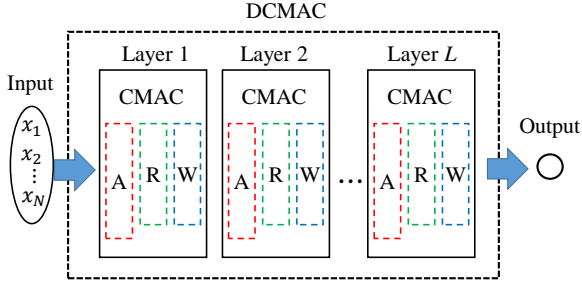

Figure 2. Architecture of the deep CMAC.

### 2.2 Review of the CMAC Model
This section reviews the structure and parameter-learning algorithm of the CMAC model.

*A. Structure of a CMAC*
Fig. 3 shows a CMAC model with five spaces: an input space, an association memory space, a receptive field space, a weight memory space, and an output space. The main functions of these five spaces are as follows:

1) Input space: This space is the input of the CMAC. In Fig. 3, the input vector is $x = [x_1, x_2, \cdots, x_N]^T \in R^N$, where $N$ is the feature dimension.

2) Association memory space: This space holds the excitation functions of the CMAC, and it has a multi-layer concept. Please note that the layers here (indicating the depth of association memory space) are different from those presented in Section 2.1 (indicating the number of CMACs in a DCMAC). To avoid confusion, we call the layer for the association memory "AS_layer" and the layer for the CMAC number "layer" in the following discussion. Fig. 4 shows an example of an association memory space with a two-dimensional input vector, $x = [x_1, x_2]^T$ with $N = 2$. The LB and UB are lower and upper bounds, respectively. We first divide $x_1$ into blocks (A, B) and $x_2$ into blocks (a, b). Next, by shifting each variable an element, we get blocks (C, D) and blocks (c, d) for the second AS_layer.

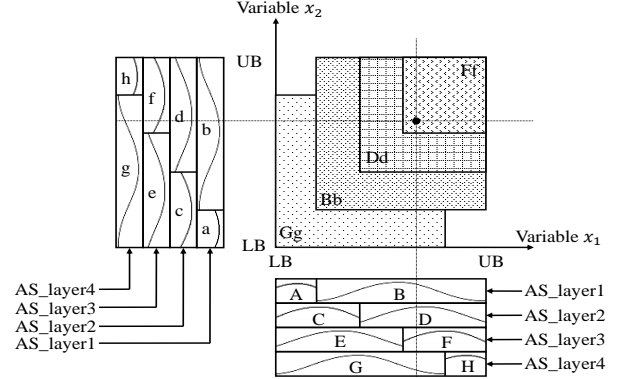

Figure 4. CMAC with a two-dimensional vector ($N = 2$).

Likewise, by shifting another variable, we can generate another AS_layer. In Fig. 4, we have four AS_layers, each AS_layer having two blocks. Therefore, the block number is eight ($N_B = 8$) for one variable; accordingly, the overall association memory space has 16 blocks ($N_A = N_B \times N$). Each block contains an excitation function, which must be a continuously bounded function, such as the Gaussian, triangular, or wavelet function. In this study, we use the Gaussian function [as shown in Fig. 4]:

$$\varphi_{ij} = \exp\left[\frac{-(x_i - m_{ij})^2}{\sigma_{ij}^2}\right], j = 1, \cdots, N_B; \ i = 1, \cdots, N, \quad (1)$$

where $x_i$ is the input signal, and $m_{ij}$ and $\sigma_{ij}$ represent the associative memory functions within the mean and variance, respectively, of the $i$-th input of the $j$-th block.

3) Receptive field space: In Fig. 4, areas formed by blocks are called receptive fields. The receptive field space has eight areas ($N_R = 8$): Aa, Bb, Cc, Dd, Ee, Ff, Gg, and Hh. Given the input $x$, the $j$-th receptive field function is represented as

$$b_j = \prod_{i=1}^N \varphi_{ij} = \exp\left[-\left(\sum_{i=1}^N \frac{(x_i - m_{ij})^2}{\sigma_{ij}^2}\right)\right]. \quad (2)$$

In the following, we express the receptive field functions in the form of vectors, namely, $\boldsymbol{b} = [b_1, b_2, \cdots, b_{N_R}]^T$.

4) Weight memory space: This space specifies the adjustable weights of the results of the receptive field space:

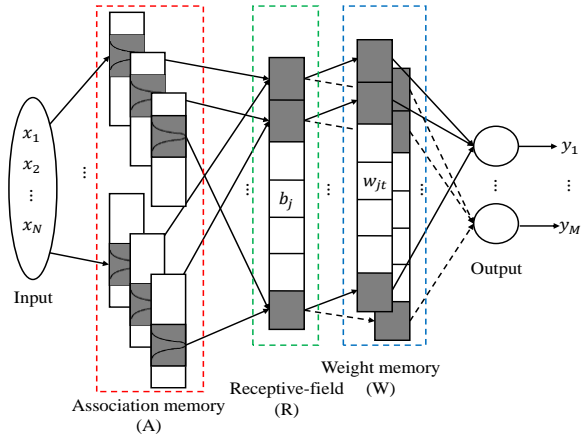

Figure 3. Architecture of a CMAC.

$$\mathbf{w}_t = [w_{1t}, w_{2t}, \cdots, w_{N_Rt}]^T \text{ for } t = 1, 2, \cdots, M, \quad (3)$$

where $M$ denotes the output vector dimension.

5) Output space: From Fig. 3, the output of the CMAC is

$$y_t = \mathbf{w}_t^T \mathbf{b} = \sum_{j=1}^{N_R} w_{jt} \exp\left[-\left(\sum_{i=1}^{N} \frac{(x_i - m_{ij})^2}{\sigma_{ij}^2}\right)\right], \quad (4)$$

where $y_t$ is the $t$-th element of the output vector, $\mathbf{y} = [y_1, y_2, \cdots, y_M]^T$. The output of state point is the algebraic sum of outputs of receptive fields (Aa, Bb, Cc, Dd, Ee, Ff, Gg, and Hh) multiplied by the corresponding weights.

### B. Parameters of Adaptive Learning Algorithm

To estimate the parameters in the association memory, receptive field, and weight memory spaces of the CMAC, we first define an objective function:

$$O(k) = \frac{1}{2} \sum_{t=1}^{M} [e_t(k)]^2, \quad (5)$$

where error signal $e_t(k) = d_t(k) - y_t(k)$, indicating the error between the desired response $d_t(k)$ and the filter's output $y_t(k)$, at the $k$-th sample. Based on Eq. (5), the normalized gradient descent method can be used to derive the update rules for the parameters in a CMAC model:

$$m_{ij}(k+1) = m_{ij}(k) + \mu_m \frac{\partial O}{\partial m_{ij}}, \quad (6)$$

where $\frac{\partial O}{\partial m_{ij}} = b_j \frac{2(x_i - m_{ij})}{(\sigma_{ij})^2} \left(\sum_{t=1}^{M} e_t w_{jt}\right);$

$$\sigma_{ij}(k+1) = \sigma_{ij}(k) + \mu_\sigma \frac{\partial O}{\partial \sigma_{ij}}, \quad (7)$$

where $\frac{\partial O}{\partial \sigma_{ij}} = b_j \frac{2(x_i - m_{ij})^2}{(\sigma_{ij})^3} \left(\sum_{t=1}^{M} e_t w_{jt}\right);$

$$w_{jt}(k+1) = w_{jt}(k) + \mu_w \frac{\partial O}{\partial w_{jt}}, \quad (8)$$

where $\frac{\partial O}{\partial w_{jt}} = e_t b_j.$

### 2.3 Proposed DCMAC Model

#### A. Structure of the DCMAC

From Eq. (4), the output of the first layer $\mathbf{y}^1$ is obtained by

$$y_t^1 = \sum_{j=1}^{N_R^1} w_{jt}^1 \exp\left[-\left(\sum_{i=1}^{N} \frac{(x_i - m_{ij}^1)^2}{\sigma_{ij}^{1\,2}}\right)\right], \quad (9)$$

where $y_t^1$ is the $t$-th element of the output of $\mathbf{y}^1$, and $N_R^1$ is the number of receptive fields in the first layer. Next, the correlation of the output of the $(l-1)$-th layer ($\mathbf{y}^{l-1}$) and that of the $l$-th layer ($\mathbf{y}^l$) can be formulated as

$$y_t^l = \sum_{j=1}^{N_R^l} w_{jt}^l \exp\left[-\left(\sum_{i=1}^{N^l} \frac{(y_i^{l-1} - m_{ij}^l)^2}{\sigma_{ij}^{l\,2}}\right)\right], l = 2 \sim L, \quad (10)$$

where $N^l$ is the input dimension of the $l$-th layer (output dimension of the $(l-1)$-th layer); $N_R^l$ is the number of receptive fields in the $l$-th layer; $y_t^l$ is the $t$-th element of $\mathbf{y}^l$; $m_{ij}^l$, $\sigma_{ij}^l$, and $\mathbf{w}_t^l$ are the parameters in the $l$-th CMAC; $L$ is the total layer number of CMAC in a DCMAC.

#### 1) Backpropagation Algorithm for DCMAC

Assume that the output vector of a DCMAC is $\mathbf{y}^L = [y_1^L, y_2^L, \cdots, y_{M^L}^L]^T \in R^{M^L}$, where $M^L$ is the feature dimension, the objective function of the DCMAC is

$$O(k) = \frac{1}{2} \sum_{t=1}^{M^L} [d_t(k) - y_t^L(k)]^2. \quad (11)$$

In the following, we present the backpropagation algorithm to estimate the parameters in the DCMAC. Because the update rules for "means and variances" and "weights" are different, they are presented separately.

1) The update algorithm of means and variances:

The update algorithms of the means and variances for the last layer (the $L$-th layer) of DCMAC are the same as that of CMAC (as shown in Eqs. (6) and (7)). For the penultimate layer (the $(L-1)$-th layer), the parameter updates are:

$$\frac{\partial O}{\partial z_{ip}^{L-1}} = \frac{\partial b_p^{L-1}}{\partial z_{ip}^{L-1}} \frac{\partial O}{\partial b_p^{L-1}}, \quad (12)$$

where $b_p^{L-1}$ is the $p$-th receptive field function of the $(L-1)$-th layer. We define the momentum $\delta_{z_p}^{L-1} = \frac{\partial O}{\partial b_p^{L-1}}$ of the $p$-th receptive field function in the $(L-1)$-th layer. Then, we have

$$\delta_{z_p}^{L-1} = \sum_{j=1}^{N_R^L} \frac{\partial b_j^L}{\partial b_p^{L-1}} \frac{\partial O}{\partial b_j^L}, \quad (13)$$

where $b_j^L$ is the $j$-th receptive field function for the $L$-th layer. Notably, by replacing $z$ with $m$ and $\sigma$ in Eq. (13), we obtain momentums $\delta_{m_p}^{L-1}$ and $\delta_{\sigma_p}^{L-1}$.

Similarly, we can derive the momentum, $\delta_{z_q}^{L-2}$, for the $q$-th receptive field function in the $(L-2)$-th layer by:

$$\delta_{z_q}^{L-2} = \frac{\partial O}{\partial b_q^{L-2}}$$
$$= \sum_{p=1}^{N_R^{L-1}} \frac{\partial b_p^{L-1}}{\partial b_q^{L-2}} \frac{\partial O}{\partial b_p^{L-1}} \quad (14)$$
$$= \sum_{p=1}^{N_R^{L-1}} \frac{\partial b_p^{L-1}}{\partial b_q^{L-2}} \delta_{z_p}^{L-1},$$

where $\partial b_q^{L-2}$ is the $q$-th receptive field function for the $(L-2)$-th layer, and $N_R^{L-1}$ is the number of receptive fields in the $(L-1)$-th layer. Based on the normalized gradient descent method, the learning of $m_{ij}^l$ (the $i$-th mean parameter in the $j$-th receptive field in the $l$-th layer) is:

$$m_{ij}^l(k+1) = m_{ij}^l(k) + \mu_{m^l} \frac{\partial b_j^l}{\partial m_{ij}^l} \delta_{m_j}^l; \quad (15)$$

similarly, the learning algorithm of $\sigma_{ij}^l$ (the $i$-th variance parameter in the $j$-th receptive field in the $l$-th layer) is:

$$\sigma_{ij}^l(k+1) = \sigma_{ij}^l(k) + \mu_{\sigma^l} \frac{\partial b_j^l}{\partial \sigma_{ij}^l} \delta_{\sigma_j}^l, \quad (16)$$

where $\mu_{m^l}$ in Eq. (15) and $\mu_{\sigma^l}$ in Eq. (16) are the learning rates for the mean and variance updates, respectively.

2) The update algorithm of weights:

The update rule of the weight in the last layer (the $L$-th layer) of DCMAC is the same as that for CMAC. For the penultimate layer (the $(L-1)$-th layer), the parameter update is:

$$\frac{\partial O}{\partial w_{jt}^{L-1}} = \frac{\partial y_t^{L-1}}{\partial w_{jt}^{L-1}} \frac{\partial O}{\partial y_t^{L-1}}, \quad (17)$$

where the momentum of the (L-1)-th layer $\delta_{w_t}^{L-1} = \frac{\partial O}{\partial y_t^{L-1}}$:

$$\delta_{w_t}^{L-1} = \sum_{j=1}^{N_R^L} \frac{\partial b_j^L}{\partial y_t^{L-1}} \sum_{r=1}^{M^L} \frac{\partial y_r^L}{\partial b_j^L} \frac{\partial O}{\partial y_r^L}, \quad (18)$$

where $y_r^L$ is the $r$-th element of the $\mathbf{y}^L$. Similarly, the momentum for the (L-2)-th layer can be computed by:

$$\begin{aligned}\delta_{w_c}^{L-2} &= \sum_{j=1}^{N_R^{L-1}} \frac{\partial b_j^{L-1}}{\partial y_c^{L-2}} \sum_{t=1}^{M^{L-1}} \frac{\partial y_t^{L-1}}{\partial b_j^{L-1}} \frac{\partial O}{\partial y_t^{L-1}} \\ &= \sum_{j=1}^{N_R^{L-1}} \frac{\partial b_j^{L-1}}{\partial y_c^{L-2}} \sum_{t=1}^{M^{L-1}} \frac{\partial y_t^{L-1}}{\partial b_j^{L-1}} \delta_{w_t}^{L-1}.\end{aligned} \quad (19)$$

According to the normalized gradient descent method, the learning algorithm of $w_{jt}^l$ (weight for the $j$-th receptive field and the $t$-th output in the $l$-th layer) is defined as

$$w_{jt}^l(k+1) = w_{jt}^l(k) + \mu_{w^l} \frac{\partial y_t^l}{\partial w_{jt}^l} \delta_{w_t}^l, \quad (20)$$

where $\mu_{w^l}$ is the learning rate for the weights.

### III. EXPERIMENTS

*3.1 Experimental Setup*

In the experiment, we consider the signal of interest $s(k) = \sin(0.06k)$ multiplied by a white noise signal, normalized within $[-1, 1]$, as shown in Fig. 5 (A). The noise signal, $n(k)$, is generated by white noise, normalized within $[-1.5, 1.5]$. A total of 1200 training samples are used in this experiment. The noise signal $n(k)$ will go through a nonlinear channel generating the damage noise $z(k)$. The relation between $n(k)$ and $z(k)$ is $z(k) = F(n(k))$, where $F(\cdot)$ represents the function of the nonlinear channel. In this experiment, we used twelve different functions, $\{ 0.6 \cdot (n(k))^{2i-1}; 0.6 \cdot \cos((n(k))^{2i-1}); 0.6 \cdot \sin((n(k))^{2i-1}), i=1, 2, 3, 4 \}$ to generate four different damage noise signals $z(k)$. The noisy signals $v(k)$ associated with four different $z(k)$ signals, with three representative channel functions, namely, $F = 0.6 \cdot (n(k))^3$, $F(\cdot) = 0.6 \cdot \cos((n(k))^3)$, and $F(\cdot) = 0.6 \cdot \sin((n(k))^3)$ are shown in Figs. 5 (B), (C), and (D), respectively.

We followed reference [8] to set up the parameters of the DCMAC, as characterized below:
1) Number of layers ($AS\_layer$): 4
2) Number of blocks ($N_B$)=8: $Ceil(5 \ (N_e)/ 4 \ (AS\_layer)) \times 4 \ (AS\_layer) = 8$.
3) Number of receptive fields ($N_R$)= 8.
4) Associative memory functions: $\varphi_{ij} = exp\left[-(x_i - m_{ij})^2/\sigma_{ij}^2\right], i = 1; \ j = 1, \cdots, N_R$.

Note that $Ceil(\cdot)$ represents the unconditional carry of the remainder. Signal range detection is required to set the UB and LB necessary to include all the signals. In this study, our preliminary results show that [UB LB]=[3 -3] gives the best performance. Please note that the main goal this study is to investigate whether DCMAC can yield better ANC results than a single-layer CMAC. Therefore, we report the results using [3 -3] for both CMAC and DCMAC in the following discussions. The initial means of the Gaussian function ($m_{ij}$) are set in the middle of each block ($N_B$) and the initial variances of the Gaussian function ($\sigma_{ij}$) are determined by the size of each block ($N_B$). With [UB LB]=[3 -3], we initialize the mean parameters as: $m_{i1} = -2.4$, $m_{i2} = -1.8$, $m_{i3} = -1.2$, $m_{i4} = -0.6$, $m_{i5} = 0.6$, $m_{i6} = 1.2$, $m_{i7} = 1.8$, $m_{i8} = 2.4$, so that the eight blocks can cover [UB LB] more evenly. Meanwhile, we set $\sigma_{ij} = 0.6$ for $j$=1,..8, and the initial weights ($w_{jt}$) zeros. Based on our experiments, the parameters initialized differently only affect the performance at the first few epochs and converge to similar values quickly. The learning rates are chosen as $\mu_s = \mu_z = \mu_w = \mu_m = \mu_\sigma = 0.001$. The parameters within all layers of the DCMAC are the same. In this study, we examine the performance of DCMACs formed by three, five, and seven layers of CMACs, which are denoted as DCMAC(3), DCMAC(5), and DCMAC(7), respectively. The input dimension was set as $N$=1; the output dimensions for CMAC and DCMACs were set as $M = 1$ and $M^L = 1$, respectively.

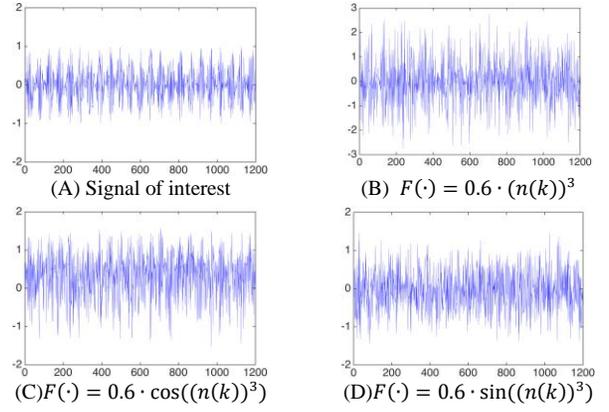

Figure 5. (A) Signal of interest $s(k)$. (B-D) Noisy signal $v(k)$ with three channel functions.

*3.2 Experimental Results*

This section compares DCMAC with different algorithms based on two performance metrics, the MSE and the convergence speed. Fig. 6 shows the converged MSE under a CMAC and a DCMAC under the three different settings, ($AS\_layer = 2$, $N_e = 5$), ($AS\_layer = 4$, $N_e = 5$), and ($AS\_layer = 4$, $N_e = 9$) testing on the channel function $F(\cdot) = 0.6 \cdot \cos((n(k))^3)$. To compare the performance of the proposed DCMAC, we conducted experiments using two popular adaptive filter methods, namely LMS [1] and the Volterra filter [5, 6]. For a fair comparison, the learning epochs are set the same for LMS, Volterra, CMAC, and DCMAC, where there are 1200 data samples in each epoch. The parameters of LMS and the Volterra filter are tested and the best results are reported in Fig. 6. From Fig. 6, we see that DCMAC outperforms not only Volterra and LMS, but also CMAC under the three setups. The same trends are observed across the 12 channel functions, and thus only the result of $F(\cdot) = 0.6 \cdot \cos((n(k))^3)$ is presented as a representative.

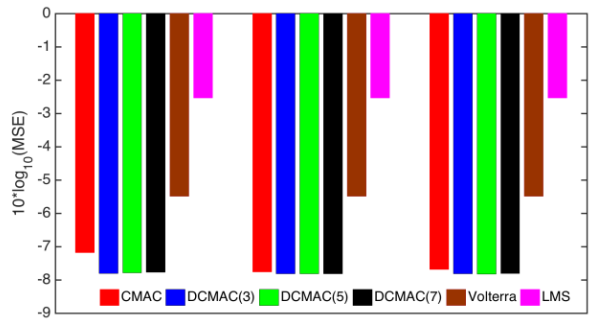

Figure 6. MSE of LMS, Volterra, CMAC, and DCMAC with channel function $F(\cdot) = 0.6 \cdot \cos((n(k))^3)$.

Speed is also an important performance metric for ANC tasks. Fig. 7 shows the convergence speed and MSE reduction rate versus number of epochs, for different algorithms. For ease of comparison and due to limited space, Fig. 7 only shows the results of three-layer DCMAC (denoted as DCMAC in Fig. 7) since the trends of DCMAC performances are consistent across different layer numbers. For CMAC and DCMAC, we adopted AS_layer = 4, $N_e = 5$. Fig. 7 shows the results of three channel functions: $F(\cdot) = 0.6 \cdot (n(k))^3$, $F(\cdot) = 0.6 \cdot \cos((n(k))^3)$, and $F(\cdot) = 0.6 \cdot \sin((n(k))^3)$. The results in Fig. 7 show that LMS and Volterra yield better performance than CMAC and DCMAC when the number of epoch is few. On the other hand, when the number of epoch increases, both DCMAC and CMAC give lower MSEs compared to that from LMS and Volterra, over all testing channels. Moreover, DCMAC consistently outperforms CMAC with lower converged MSE scores. The results show that the performance gain of the DCMAC becomes increasingly more significant as the nonlinearity of the channels increases. Finally, we note that the performances of both DCMAC and CMAC became saturated around 400 epochs. In a real-world application, a development set of data can be used to determine the saturation point, so that the adaptation can be switched off.

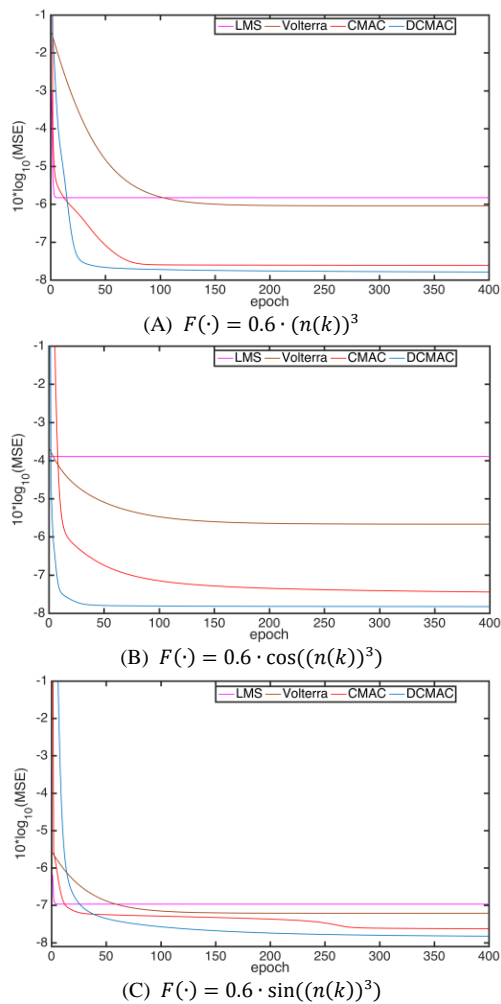

(A) $F(\cdot) = 0.6 \cdot (n(k))^3$

(B) $F(\cdot) = 0.6 \cdot \cos((n(k))^3)$

(C) $F(\cdot) = 0.6 \cdot \sin((n(k))^3)$

Figure 7. MSEs of LMS, Volterra, CMAC, and DCMAC with three types of channel functions. More results are presented in https://goo.gl/frAZvk

Simulation results of a CMAC and that of a DCMAC, both for 400 epochs of training, are shown in Figs. 8 (A) and (B), respectively. The results show that the proposed DCMAC can achieve better filtering performance than that from the CMAC for this noise cancellation system.

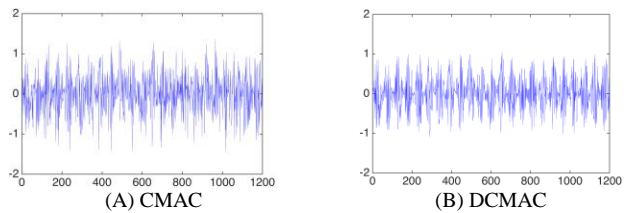

(A) CMAC        (B) DCMAC

Figure 8. Recovered signal using (A) CMAC and (B) DCMAC, where $F(\cdot) = 0.6 \cdot \cos((n(k))^3)$.

Table I lists the mean and variance of MSE scores for LMS, Volterra, CMAC, and DCMAC across 12 channel functions. The MSE for each method at a channel function was obtained with 1000 epochs of training. From the results, both CMAC and DCMAC give lower MSE than LMS and Volterra. In addition to the results in Table I, we adopted the dependent t-Test for the hypothesis test on the 12 sets of results. The t-Test results revealed that DCMAC outperforms CMAC with $P$-values = 0.005.

TABLE I.
MEAN AND VAIRAINCE OF MSES FOR LMS, VOLTERRA, CMAC, AND DCMAC OVER 12 CHANNEL FUCNTIONS

|          | LMS   | Volterra | CMAC  | DCMAC |
|----------|-------|----------|-------|-------|
| Mean     | −4.35 | −5.05    | −7.01 | −7.59 |
| Variance | 11.95 | 11.57    | 1.08  | 0.19  |

IV. CONCLUSION

The contribution of the present study was two-fold: First, inspired by the recent success of deep learning algorithms, we extended the CMAC structure into a deep one, termed deep CMAC (DCMAC). Second, a backpropagation algorithm was derived to estimate the DCMAC parameters. Due to the five-space structure, the backpropagation for DCAMC is different from that used in the related artificial neural networks. The parameter updates involved in DCMAC training include two parts (1) The update algorithm of means and variances; (2) The update algorithm of weights. Experimental results of the ANC tasks showed that the proposed DCMAC can achieve better noise cancellation performance when compared with that from the conventional single-layer CMAC. In future, we will investigate the capabilities of the DCMAC on other signal processing tasks, such as echo cancellation and single-microphone noise reduction. Meanwhile, advanced deep learning algorithms used in deep neural networks, such as dropout and sparsity constraints, will be included in the DCMAC framework. Finally, like related deep learning researches, identifying a way to optimize the number of layers and initial parameters in DCMAC per the amount of training data are important future works.